\documentclass[twoside]{article}

\usepackage{graphicx}

\usepackage{subeqn} 
\usepackage{subfigure}

\usepackage[T2A]{fontenc}
\usepackage[cp1251]{inputenc}
\usepackage{cmpj2e}

\title[The DSUB$m$ Approximation Scheme for the Coupled Cluster
Method]%
{The DSUB$m$ Approximation Scheme for the Coupled Cluster Method and
  Applications to Quantum Magnets}
\author[R.F. Bishop {\it et al.}]{R.F. Bishop\refaddr{label1},
  P.H.Y. Li\refaddr{label1}, J. Schulenburg\refaddr{label2}}
\addresses{ 
\addr{label1} School of Physics and Astronomy, Schuster
  Building, The University of Manchester, Manchester, M13 9PL, UK
\addr{label2} Universit\"{a}tsrechenzentrum, Universit\"{a}t
  Magdeburg, P.O. Box 4120, 39016 Magdeburg, Germany }
\begin{document}

\maketitle

\begin{abstract}
  A new approximate scheme, DSUB$m$, is described for the coupled
  cluster method.  We then apply it to two well-studied (spin-1/2
  Heisenberg antiferromagnet) spin-lattice models, namely: the $XXZ$
  and the $XY$ models on the square lattice in two dimensions.
  Results are obtained in each case for the ground-state energy, the
  sublattice magnetization and the quantum critical point.  They are in
  good agreement with those from such alternative methods as spin-wave theory, series
  expansions, quantum Monte Carlo methods and those from the CCM using
  the LSUB$m$ scheme.
\pacs 75.10.Jm, 75.30.Gw, 75.40.-s, 75.50.Ee  
\end{abstract}

\section{Introduction}           
The coupled cluster method (CCM) is a universal microscopic technique
of quantum many-body
theory~\cite{Co:1958,Ci:1966,Pa:1972,Ku:1978,Ar:1983,Ar:1987,Ba:1989,Bi:1991,Bi:1998_LectPhys_V510}.
It has been applied successfully to many physical systems including:
\begin{itemize}    
\item systems existing in the spatial continuum, e.g., the electron
  gas~\cite{Bi:1978,Bi:1982}, atomic nuclei and nuclear matter~\cite{Da:1981,Da:1981_b},
  and molecules~\cite{Ba:1981}; and
\item systems on a discrete spatial lattice, e.g., spin-lattice
  models of quantum
  magnetism~\cite{Ro:1990,Bi:1991_b,Bu:1995,Fa:1997,Bi:1998,Bi:2000,Fa:2001,Fa:2002,Bi:2008_JPCM_V20_p255252,Bi:2008_EPL,Bi:2008_PRB,Bi:2008_JPCM_V20_p415213,Bi:2009_PRB79}.
\end{itemize}
A special characteristic of the CCM is that it deals with infinite
systems from the outset, and hence one never needs to take the limit
$N \rightarrow \infty$ explicitly in the number, $N$, of interacting
particles or the number of sites in the lattice. However,
approximations in the inherent cluster expansions for the correlation
operators are required.  Several efficient and systematic
approximation schemes for the CCM have been specifically developed by
us previously for use with lattice
systems~\cite{Bi:1991_b,Bi:1991_c,Bi:1994,Fa:2004}.  Up till now the
most favoured and most successful CCM approximation schemes for
lattices have been the so-called LSUB$m$ and SUB$n$--$m$ schemes that
we describe more fully below in section~\ref{spin_latt}.  Although the
LSUB$m$ scheme, in particular, has been shown to be highly successful
in practice for a wide variety of both frustrated and unfrustrated
spin-lattice systems, a disadvantage of this scheme is that the number
of spin configurations generally rises very rapidly with the
truncation index $m$. This motivates us to develop alternative schemes
which satisfy one or both of the following two criteria:
\begin{itemize}
\item that we are able to calculate more levels of approximation
  within the scheme, and hence have available more data points for the
  necessary extrapolations for calculated physical quantities to the
  exact limit where all configurations are retained; and
\item that one can capture the physically most important
  configurations at relatively low orders, so that the quantities of
  interest converge more rapidly with the truncation index.
\end{itemize}
                    
A main aim of our work is thus to provide users of the CCM with more
choices of approximation scheme. In this paper we outline the formal
aspects, of and present preliminary results for some benchmark models
for, a new CCM approximation scheme, called the DSUB$m$ scheme, for use
with systems described on a spatial lattice.  The rest of the paper is
organized as follows. In section~\ref{ccm_formalism} we introduce the
CCM formalism in general. In section~\ref{spin_latt} we then discuss
the specific application of the CCM to spin-lattice systems. We
consider in section~\ref{ApproxSchm} both the existing truncation
schemes and introduce the alternative new DSUB$m$ scheme.  In order to
evaluate the accuracy of the new approximation scheme we apply it to
two very well-studied antiferromagnetic spin-lattice
models~\cite{Fa:1997,Bi:2000}, namely: the $XXZ$ and the $XY$ models
on the square lattice in two dimensions (2D). Both models have a
quantum phase transition. They have also been very successfully
investigated by the CCM within the LSUB$m$ scheme for the ground-state
(gs) energy and the gs order parameter which is, in our case, the
sublattice magnetization. All techniques applied to lattice spin
systems need to be extrapolated in terms of some appropriate
parameters. For exact diagonalization and quantum Monte Carlo methods,
this is the lattice size $N$. As noted above, one good aspect of the
CCM is that we may work in the limit of infinite lattice size ($N
\rightarrow \infty$) from the outset. By contrast, the extrapolation
for the CCM is in terms of some truncation index $m$, where in the
limit $m \rightarrow \infty$ we retain {\it all} configurations. The
CCM extrapolations that have been used up till now in the trunction
index, e.g., $m$ for the LSUB$m$ approximation, are heuristic schemes,
but we have considerable prior
experience~\cite{Bi:2000,Bi:1994,Ze:1998,Kr:2000,Schm:2006} in using
and refining them, as described in
section~\ref{Extrapo}. The new DSUB$m$ approximation scheme is then
applied to the square-lattice spin-$1/2$ antiferromagnetic $XXZ$ model
in section~\ref{XXZ} and the corresondingly $XY$ model in section~\ref{XY},
respectively. The results for both models are compared critically with
those from the corresponding LSUB$m$ scheme and also with the results
from other methods. Finally, our conclusions are given in
section~\ref{discussion} where we reiterate a brief summary of the
results.

\section{The CCM Formalism}
\label{ccm_formalism}
This section briefly describes the CCM formalism (and see e.g.,
Refs.~\cite{Bi:1991,Bi:1998_LectPhys_V510} for further details). A
first step in any CCM application is to choose a normzalized model (or
reference) state $|\Phi\rangle$ that can act as a cyclic vector with
respect to a complete set of mutually commuting multiconfigurational
creation operators $C^{+}_{I} \equiv (C^{-}_{I})^{\dagger}$. The index
$I$ here is a set-index that labels the many-particle configuration
created in the state $C^{+}_{I}|\Phi\rangle$. The exact ket and bra gs
energy eigenstates $|\Psi\rangle$ and $\langle\tilde{\Psi}|$, of the
many-body system are then parametrised in the CCM form as:
\vskip0.2cm
\hspace{1.7in} \textbf{Ket-state} \qquad  \textbf{Bra-state} 
\begin{equation}
|\Psi\rangle = \mbox{e}^{S}|\Phi\rangle \,; \qquad \langle\tilde{\Psi}| = \langle\Phi|\tilde{S}\mbox{e}^{-S} \,,   \label{eq:ket_bra_eq}
\end{equation}
\begin{equation}
S = \sum_{I\neq0}{\cal S}_{I}C^{+}_{I}\,; \qquad \tilde{S} = 1 + \sum_{I\neq0}\tilde{{\cal S}_{I}}C^{-}_{I}\,,   \label{eq:ket_bra_coefficients}
\end{equation}
\begin{equation}
H|\Psi\rangle = E|\Psi\rangle \,; \qquad \langle\tilde{\Psi}|H = E\langle\tilde{\Psi}| \,,  \label{eq:SE_CCM}
\end{equation}
where we have defined $C^{+}_{0} \equiv 1 \equiv C^{-}_{0}$. The
requirements on the multiconfigurational creation operators are that
any many-particle state can be written exactly and uniquely as a
linear combination of the states $\{C^{+}_{I}|\Phi\rangle\}$, which
hence fulfill the completeness relation
\begin{equation}
\sum_{I}C^{+}_{I}|\Phi\rangle \langle\Phi|C^{-}_{I} = 1 = |\Phi\rangle \langle\Phi| + \sum_{I \neq 0}C^{+}_{I}|\Phi\rangle \langle\Phi|C^{-}_{I}\, ,
\end{equation}
together with the conditions,
\begin{equation}
C^{-}_{I}|\Phi\rangle = 0 = \langle\Phi| C^{+}_{I}\,; \qquad \forall \emph{I} \neq 0 \, ,
\end{equation}
\begin{equation}
[C^{+}_{I},C^{+}_{J}]=0=[C^{-}_{I},C^{-}_{J}]\, .  \label{commutation}
\end{equation}

Approximations are necessary in practice to restrict the label set
{{\em I}} to some finite (e\@.g\@., LSUB$m$) or infinite (e.g.,
SUB$n$) subset, as described more fully below. The correlation operator
$S$ is a linked-cluster operator and is decomposed in terms of a
complete set of creation operators ${C^{+}_{I}}$. It creates
excitations on the model state by acting on it to produce correlated
cluster states. Although the manifest Hermiticity,
$(\langle\tilde{\Psi}|)^{\dagger}\equiv|\Psi\rangle/\langle\Psi|\Psi\rangle$,
is lost, the normalization conditions
$\langle\tilde{\Psi}|\Psi\rangle=\langle\Phi|\Psi\rangle=\langle\Phi|\Phi\rangle\equiv
1$ are preserved. The CCM Schr\"{o}dinger equations (\ref{eq:SE_CCM})
are thus writtern as
\begin{equation}
H \mbox{e}^{S}|\Phi\rangle = E \mbox{e}^{S}|\Phi\rangle\,; \qquad \langle\Phi|\tilde{S}\mbox{e}^{-S}H=E\langle\Phi|\tilde{S}\mbox{e}^{-S}\,;  \label{eq:CCM_SE}
\end{equation}
and its equivalent similarity-transformed form becomes
\begin{equation}
\mbox{e}^{-S}H\mbox{e}^{S}|\Phi\rangle = E|\Phi\rangle \,;\qquad \langle\Phi|\tilde{S}\mbox{e}^{-S}H\mbox{e}^{S} = E\langle\Phi|\tilde{S}\,.  \label{eq:SE_similarity_trans}
\end{equation}

We note that while the parametrizations of equations
(\ref{eq:ket_bra_eq}) and (\ref{eq:ket_bra_coefficients}) are not
manifestly Hermitian conjugate, they do preserve the important
Hellmann-Feynman theorem at all levels of approximations (viz., when
the complete set of many-particle configurations {$I$} is
truncated)~\cite{Bi:1998_LectPhys_V510}. Furthermore, the amplitudes
(${\cal S}_{I}, \tilde{\cal S}_{I}$) form canonically conjugate pairs
in a time-dependent version of the CCM, by contrast with the pairs
(${\cal S}_{I}, {\cal S}^{\ast}_{I}$) coming from a manifestly
Hermitian-conjugate representation for
$\langle\tilde{\Psi}|=(\langle\Phi|\mbox{e}^{S^{\dagger}}\mbox{e}^{S}|\Phi\rangle)^{-1}\langle\Phi|\mbox{e}^{S^{\dagger}}$,
which are {\it not} canonically conjugate to one another.

The static gs CCM correlation operators $S$ and $\tilde{S}$ contain
the real $c$-number correlation coefficients ${\cal S}_{I}$ and
${\tilde{\cal S}}_{I}$ that need to be calculated. Clearly, once the
coefficients $\{{\cal S}_{I}, \tilde{\cal S}_{I}\}$ are known, all
other gs properties of the many-body system can be derived from
them. For example, the gs expectation value of an arbitrary operator
$A$ can be expressed as
\begin{equation}
\bar{A} \equiv \langle A \rangle \equiv \langle\tilde{\Psi}|A|\Psi\rangle = \langle\Phi|\tilde{S}\mbox{e}^{-S}A\mbox{e}^{S}|\Phi\rangle \equiv A({\cal S}_{I}, \tilde{{\cal S}_{I}})\,.   \label{bar_A}
\end{equation}

To find the gs correlation coefficients $\{{\cal S}_{I}, \tilde{\cal
  S}_{I}\}$ we simply insert the parametrization of equation
(\ref{eq:ket_bra_coefficients}) into the Schr\"{o}dinger equations
(\ref{eq:SE_similarity_trans}), and project onto the complete sets of
states $\{\langle\Phi|C^{-}_{I}\}$ and $\{C^{+}_{I}|\Phi\rangle\}$,
respectively,
\begin{equation}
\langle \Phi|C^{-}_{I}\mbox{e}^{-S}H\mbox{e}^{S}|\Phi\rangle = 0\,; \qquad  \forall I \neq 0\,.    \label{eq:ket_coeff}
\end{equation}
\begin{equation}
 \langle\Phi|\tilde{S}(\mbox{e}^{-S}H\mbox{e}^{S} - E)C^{+}_{I}|\Phi\rangle = 0\;;\hspace{.2in} \forall I \neq 0\,.  \label{eq:Bra_coeff_1}
\end{equation}
Equation (\ref{eq:Bra_coeff_1}) may also easily be rewritten, by
pre-multiplying the ket-state equation (\ref{eq:SE_similarity_trans})
with the state $\langle\Phi|\tilde{S}C^{+}_{I}$ and using the
commutation relation (\ref{commutation}), in the form
\begin{equation}
\langle\Phi|\tilde{S}\mbox{e}^{-S}[H, C^{+}_{I}]\mbox{e}^{S}|\Phi\rangle = 0\,; \qquad \forall I \neq 0 \,.  \label{eq:Bra_coeff_2}
\end{equation}  
Equations (\ref{eq:ket_coeff})--(\ref{eq:Bra_coeff_2}) may be
equivalently derived by requiring that the gs energy expectation
value,
$\bar{H}\equiv\langle\tilde{\Psi}|H|\Psi\rangle=\langle\Phi|\tilde{S}\mbox{e}^{-S}H\mbox{e}^{S}|\Phi\rangle$,
is minimized with respect to the entire set $\{{\cal
  S}_{I},{\tilde{\cal S}}_{I}\}$. In practice we thus need to solve
equations (\ref{eq:ket_coeff}) and (\ref{eq:Bra_coeff_2}) for the set
$\{{\cal S}_{I},{\tilde{\cal S}}_{I}\}$. We note that equations
(\ref{bar_A}) and (\ref{eq:ket_coeff}) show that the gs energy at the
stationary point has the simple form
\begin{equation}
E \equiv E({\cal S}_{I})=\langle \Phi|\mbox{e}^{-S}H\mbox{e}^{S}|\Phi\rangle\,,  \label{E}  
\end{equation}
which also follows immediately from the ket-state equation
(\ref{eq:SE_similarity_trans}) by projecting it onto the state
$\langle\Phi|$. It is important to note that this (bi-)variational
formulation does not necessarily lead to an upper bound for $E$ when
the summations over the index set $\{I\}$ for $S$ and $\tilde{S}$ in
equation (\ref{eq:ket_bra_coefficients}) are truncated, due to the
lack of manifest Hermiticity when such approximations are
made. Nevertheless, as we have pointed out above, one can
prove~\cite{Bi:1998_LectPhys_V510} that the important Hellmann-Feynman
theorem {\it is} preserved in all such approximations.

We note that equation (\ref{eq:ket_coeff}) represents a coupled set of
multinomial equations for the $c$-number correlation coefficients
$\{{\cal S}_{I}\}$. The nested commutator expansion of the
similarity-transformed Hamiltonian,
\begin{equation}
\mbox{e}^{-S}H\mbox{e}^{S} = H + [H,S] + \frac{1}{2!}[[H,S],S] + \cdots \,,  \label{eq:H_Sim_xform}
\end{equation}
and the fact that all of the individual components of $S$ in the
decomposition of equation (\ref{eq:ket_bra_coefficients}) commute with
one another by construction [and see equation (\ref{commutation})],
together imply that each element of $S$ in equation
(\ref{eq:ket_bra_coefficients}) is linked directly to the Hamiltonian
in each of the terms in equation (\ref{eq:H_Sim_xform}). Thus, each of
the coupled equations (\ref{eq:ket_coeff}) is of Goldstone {\it
  linked-cluster} type. In turn this guarantees that all extensive
variables, such as the energy, scale linearly with particle number
$N$. Thus, at any level of approximation obtained by truncation in the
summations on the index $I$ in the parametrizations of equation
(\ref{eq:ket_bra_coefficients}), we may (and, in practice, do) work
from the outset in the limit $N \rightarrow \infty$ of an infinite
system.

Furthermore, each of the seemingly infinite-order (in $S$)
linked-cluster equations (\ref{eq:ket_coeff}) will actually be of
finite length when expanded using equation (\ref{eq:H_Sim_xform}),
since the otherwise infinite series of equation (\ref{eq:H_Sim_xform})
will actually terminate at a finite order, provided only (as is
usually the case, including those for the Hamiltonians considered in
this paper) that each term in the Hamiltonian $H$ contains a finite
number of single-particle destruction operators defined with respect
to the reference (or generalized vacuum) state $|\Phi\rangle$. Hence,
the CCM parametrization naturally leads to a workable scheme that can
be implemented computationally in an efficient manner to evaluate the
set of configuration coefficients $\{{\cal S}_{I},{\tilde{\cal S}}_{I}\}$
by solving the coupled sets of equations (\ref{eq:ket_coeff}) and
(\ref{eq:Bra_coeff_2}), once we have devised practical and systematic
truncation hierarchies for limiting the set of multiconfigurational
set-indices $\{I\}$ to some suitable finite or infinite subset.

\section{CCM for Spin-Lattice Systems}
\label{spin_latt}
We will discuss various practical CCM truncation schemes that fulfill
the criteria of being systematically improvable in some suitable
truncation index $m$, and that can be extrapolated accurately in
practice to the exact, $m \rightarrow \infty$, limit for calculated physical quantities. Before doing so,
however, we first describe how the general CCM formalism described in
section~\ref{ccm_formalism} is implemented for spin-lattice problems in
practice. As is the case for {\it any} application of the CCM to a
general quantum many-body system, the first step is to choose a
suitable reference state $|\Phi\rangle$ in which the the state of the spin (viz.,
in practice, its projection onto a specific quantization axis in spin
space) on every lattice site $k$ is
characterized. Clearly, the choice of $|\Phi\rangle$ will depend on both
the system being studied and, more importantly, which of its possible phases is
being considered. We describe examples of such choices later for the
particular models that we utilize here as test cases for our new
truncation scheme.

Whatever the choice of $|\Phi\rangle$ we note firstly that is very
convenient, in order to set up as universal a methodology as possible,
to treat the spins on every lattice site in an arbitrarily given model
state $|\Phi\rangle$ as being equivalent. A suitably simple way of so doing
is to introduce a different local quantization axis and a
correspondingly different set of spin coordinates on each lattice site
$k$, so that all spins, whatever their original orientation in
$|\Phi\rangle$ in a global spin-coordinate system, align along the
same direction (which, to be specific, we henceforth choose as the
negative $z$ direction) in these local spin-coordinate frames. This
can always be done in practice by defining a suitable rotation in spin
space of the global spin coordinates at each lattice site $k$. Such
rotations are canonical transformations that leave unchanged the
fundamental spin commutation relations,
\begin{equation}
[s^{+}_{k},s^{-}_{k'}]=2s^{z}_{k}\delta_{kk'} \,; \qquad [s^{z}_{k},s^{\pm}_{k'}]=\pm s^{\pm}_{k}\delta_{kk'}\,,   \label{commutation_Sz_S+_comm_S+S-}
\end{equation}
\begin{equation}
s^{\pm}_{k}\equiv s^{x}_{k}\pm is^{y}_{k}\,,   \label{spin_rasing_lowering_operators}
\end{equation}
among the usual SU(2) spin operators ($s^{x}_{k},s^{y}_{k},s^{z}_{k}$)
on lattice site $k$. Each spin has a total spin quantum number,
$s_{k}$, where ${\bf s}^{2}_{k}=s_{k}(s_{k}+1)$ is the SU(2) Casimir
operator. For the models considered here, $s_{k}=s=1/2$, $\forall k$.

In the local spin frames defined above the configuration indices $I$
simply become a set of lattice site indices, $I \rightarrow
(k_{1},k_{2},\cdots,k_{m}$). The corresponding generalized
multiconfigurational creation operators $C^{+}_{I}$ thus become simple
products of single spin-raising operators, $C^{+}_{I} \rightarrow
s^{+}_{k_{1}}s^{+}_{k_{2}}\cdots s^{+}_{k_{m}}$, and, for example, the
ket-state CCM correlation operator is expressed as
\begin{equation}
S = \sum_{m} \sum_{{k_{1}}{k_{2}}\cdots{k_{m}}} {\cal S}_{k_{1}k_{2}\cdots k_{m}}
s^{+}_{k_{1}} s^{+}_{k_{2}} \cdots s^{+}_{k_{m}}\,,  \label{ket_operator}
\end{equation}
and $\tilde{S}$ is similarly defined in terms of the spin-lowering
operators $s^{-}_{k}$. Since the operator $S$ acts on the state
$|\Phi\rangle$, in which all spins point along the negative $z$-axis
in the local spin-coordinate frames, every lattice site $k_{i}$ in
equation (\ref{ket_operator}) can be repeated up to no more than 2$s$
times in each term where it is allowed, since a spin $s$ has only
($2s+1$) possible projections along the quantization axis.

In practice the allowed configurations are often further constrained
by symmetries in the problem and by conservation laws. An example of
the latter is provided by the $XXZ$ model considered below in
section~\ref{XXZ}, for which we can easily show that the total
$z$-component of spin, $s^{T}_{z}=\sum^{N}_{k=1}s^{z}_{k}$, in the
original global spin coordinates, is a good quantum number since
$[s^{z}_{T},H]=0$ in this case. Finally, for the quasiclassical
magnetically ordered states that we calculate here for both models in
sections~\ref{XXZ} and \ref{XY}, the order parameter is the sublattice
magnetization, $M$, which is given within the {\it local} spin
coordinates defined above as
\begin{equation}
M \equiv -\frac{1}{N}
\langle\tilde{\Psi}|\sum_{k}^{N}s^{z}_{k}|\Psi\rangle =
-\frac{1}{N}\sum_{k}^{N} \langle\Phi|\tilde{S}
\mbox{e}^{-S}s^{z}_{k}\mbox{e}^{S}|\Phi\rangle \,.   \label{M}
\end{equation}

After the local spin axes have been chosen, the model state thus has
all spins pointing downwards (i.e., in the negative $z$-direction,
where $z$ is the quantization axis),
\begin{equation}
|\Phi\rangle = \bigotimes^{N}_{k_{1}}|\downarrow\rangle_{k}\,; \qquad \mbox{in the local spin axes,}   \label{local_quan}
\end{equation}
where $|\downarrow\rangle\equiv |s,-s\rangle$ in the usual
$|s,m_{s}\rangle$ notation for single spin states.

The similarity-transformed Hamiltonian $\bar{H}\equiv
\mbox{e}^{-S}H\mbox{e}^{-S}$, and all of the corresponding matrix
elements in equations (\ref{bar_A})--(\ref{E}) and equation (\ref{M}),
for example, may then be evaluated in the local spin coordinate frames
by using the nested commutator expansion of equation
(\ref{eq:H_Sim_xform}), the commutator relations of equation
(\ref{commutation_Sz_S+_comm_S+S-}), and the simple universal
relations
\begin{equation}
s^{-}_{k}|\Phi\rangle=0\,; \qquad \forall k\,,
\end{equation}
\begin{equation}
s^{z}_{k}|\Phi\rangle\,; \qquad \forall k\,,
\end{equation}
that hold at all lattice sites in the local spin frames.

\section{Approximation schemes}
\label{ApproxSchm}
The CCM formalism is exact if all many-body configurations $I$ are
included in the $S$ and $\tilde{S}$ operators. In practice, it is
necessary to use approximation schemes to truncate the correlation
operators.

\subsection{Common previous truncation schemes}
The main approximation scheme used to date for continuous systems is
the so-called SUB$n$ scheme described below. For systems defined on a
regular periodic spatial lattice, we have a further set of
approximation schemes which are based on the discrete nature of the
lattice, such as the SUB$n$--$m$ and LSUB$n$ schemes described below.
The various schemes and their definitions for spin-lattice systems are:
\begin{enumerate}
\item the SUB$n$ scheme, in which only the correlations involving $n$
  or fewer spin-raising operators for $S$ are retained, but with no
  further restrictions on the spatial separations of the spins involved in the configurations;
\item the SUB$n$--$m$ scheme wich includes only the subset of all
  $n$-spin-flip configurations in the SUB$n$ scheme that are defined
  over all lattice animals of size $\leq m$, where a lattice animal is
  defined as a set of contiguous lattice sites, each of which is
  nearest-neighbour to at least one other in the set; and
\item the LSUB$m$ scheme which includes all possible multi-spin-flip
  configurations defined over all lattice animals of size $\leq
  m$. The LSUB$m$ scheme is thus equivalent to the SUB$n$--$m$ scheme
  for $n=2sm$ for particles of spin quantum number $s$. For example,
  for spin-1/2 systems, for which no more than one spin-raising operator,
  $s^{+}_{k}$, can be applied at each site $k$, LSUB$m \equiv$
  SUB$m$--$m$.
\end{enumerate}

\subsection{The new DSUB$m$ scheme}
Our new DSUB$m$ scheme is now defined to include in the correlation operator $S$ all possible
configurations of spins involving spin-raising operators where the
maximum length or distance of any two spins apart is defined by
$L_{m}$, where $L_{m}$ is a vector joining sites on the lattice and
the index $m$ labels lattice vectors in order of size.  Hence DSUB$1$
includes only nearest-neighbours, etc.

Table~\ref{form_Lm} shows how $L_{m}$ progresses in terms of $k$ and
$l$ (which are the sides of the lattice in the $x$ and $y$ directions)
for the case of a 2D square lattice with sides parallel to the $x$ and
$y$ axes.
\begin{table}[!b]
  \caption{The formulation of the length parameter $L_{m}$ of the DSUB$m$ approximation on a square lattice, in terms of lattice increments $k$ and $l$ along the two sides of the lattice.}   
\label{form_Lm}
\vspace{2ex}
\begin{center}
\begin{tabular}{|c|c|c|c|}  \hline\hline
DSUB$m$ & $L_{m}$ & $k$ & $l$  \\ \hline
DSUB$1$ & 1 & 0 & 1    \\ 
DSUB$2$ & $\sqrt{2}$ & 1 & 1 \\ 
DSUB$3$ & 2 & 0 & 2   \\ 
DSUB$4$ & $\sqrt{5}$ & 1 & 2 \\ 
DSUB$5$ & $\sqrt{8}$ &2 & 2  \\ 
DSUB$6$ & 3 & 0 & 3 \\ 
DSUB$7$ & $\sqrt{10}$& 1 & 3  \\ \hline\hline
\end{tabular}
\end{center}   
\end{table}
Similar tables can be constructed for an arbitrary regular lattice in
any number of dimensions. Table~\ref{form_Lm} shows, for example, that
the DSUB5 approximation on a 2D square lattice involves all clusters
of spins (and their associated spin-raising operators) for which the
maximum distance between any two spins is $\sqrt{8}$. Clearly
the DSUB$m$ scheme orders the multispin configurations in terms,
roughly, of their compactness, whereas the LSUB$m$ scheme orders them
according to the overall size of the lattice animals (or polyominos),
defined as the number of contiguous lattice sites involved.

\section{Extrapolation schemes}
\label{Extrapo}
Any of the above truncated approximations clearly becomes exact when
all possible multispin cluster configurations are retained, i.e., in
the limit as $n \rightarrow \infty$ and/or $m \rightarrow \infty$. We
have considerable experience, for example, with the appropriate
extrapolations for the LSUB$m$
scheme~\cite{Bi:2000,Bi:1994,Ze:1998,Kr:2000,Schm:2006}, that shows
that the gs energy behaves in the large-$m$ limit as a power series in
$1/m^{2}$, whereas the order parameter $M$ behaves as a power series
in $1/m$ (at least for relatively unfrustrated systems). Initial
experience with the new DSUB$m$ scheme shows, perhaps not
unsurprisingly, that in the corresponding large $m$ limit the gs
energy and order parameter behave as power series in $1/L^{2}_{m}$ and
$1/L_{m}$, respectively, as we show in more detail (below) for the two
examples of the spin-1/2 $XXZ$ and $XY$ models on the 2D square
lattice. It is clear on physical grounds that the index $L_{m}$ should
provide a better extrapolation variable than the index $m$ itself for
the DSUB$m$ scheme, and so it turns out in practice. For the present,
where we are interested primarily in a preliminary investigation of
the power and accuracy of the DSUB$m$ scheme, we limit ourselves to
retaining only the leading terms in the power-series expansions,
\begin{equation}
\frac{E}{N}\biggl\vert_{\mbox{\scriptsize{DSUB}}m}=a_{0}+a_{1}\left(\frac{1}{L_{m}^{2}}\right)\,;   \label{Extrapo_E}
\end{equation}
\begin{equation}
M\bigl\vert_{\mbox{\scriptsize{DSUB}}m}=b_{0}+b_{1}\left(\frac{1}{L_{m}}\right)\,.         \label{Extrapo_M}
\end{equation}
Further sub-leading terms in each of the power series can easily be
retained later should it prove useful to increase the accuracy of the
extrapolations.

\subsection{Three fundamental rules for the selection and extrapolation of the CCM raw data}
We list below three fundamental rules as guidelines for the selection
and extrapolation of the CCM raw data, using {\it any}
approximation scheme:
\begin{itemize}
\item RULE 1: In order to fit well to any fitting formula that
  contains $n$ unknown parameters, one should have at least ($n+1$)
  data points. This rule takes precedence over all other rules, and is
  vital to obtain a robust and stable fit.
\item RULE 2: Avoid using the lowest data points (e.g., LSUB2, SUB2-2,
  DSUB1, etc.) wherever possible, since these points are rather far
  from the large-$m$ limit, unless it is necessary to do so to avoid
  breaking RULE 1, e.g., when only $n$ data points are available.
\item RULE 3: If RULE 2 has been broken (e.g., by including LSUB2 or
  SUB$2$--$2$ data points), then do some other careful consistency
  checks on the robustness and accuracy of the results.
\end{itemize} 

\section{The spin-$1/2$ antiferromagnetic $XXZ$ model on the square lattice}
\label{XXZ}
In this section, we shall consider the spin-1/2 $XXZ$ model on the
infinite square lattice.  The Hamiltonian of the $XXZ$ model, in
global spin coordinates, is written as
\begin{equation}
H_{XXZ} = \sum_{\langle i,j \rangle}[s^{x}_{i}s^{x}_{j} + s^{y}_{i}s^{y}_{j}
+ \Delta s^{z}_{i}s^{z}_{j}]\;, \label{eq:H_XXZ}
\end{equation}
where the sum on $\langle i,j \rangle$ runs over all nearest-neighbour
pairs of sites on the lattice and counts each pair only once. Since
the square lattice is bipartite, we consider $N$ to be even, so that
each sublattice contains $(1/2)N$ spins, and we consider only the case
where $N \rightarrow \infty$. The N\'{e}el state is the ground state
(GS) in the trivial Ising limit $\Delta \rightarrow \infty$, and a
phase transition occurs at (or near to) $\Delta = 1$. Indeed, the
classical GS demonstrates perfect N\'{e}el order in the $z$-direction
for $\Delta > 1$, and a similar perfectly ordered $x$-$y$ planar
N\'{e}el phase for $-1 < \Delta < 1$. For $\Delta < -1$ the classical
GS is a ferromagnet.

The case $\Delta = 1$ is equivalent to the isotropic Heisenberg model,
whereas $\Delta = 0$ is equivalent to the isotropic version of the
$XY$ model considered in section~\ref{XY} below. The $z$ component of total
spin, $s^{z}_{T}$, is a good quantum number as it commutes with the
Hamiltonian of equation (\ref{eq:H_XXZ}). Thus one may readily check
that $[s^{z}_{T},H_{XXZ}]=0$. Our interest here is in those values of
$\Delta$ for which the GS is an antiferromagnet.

The CCM treatment of any spin system is started by choosing an
appropriate model state $|\Phi\rangle$ (for a particular regime), so
that a linear combinations of products of spin-raising operators can
be applied to this state and all possible spin configurations are
determined. There is never a unique choice of model state
$|\Phi\rangle$. Our choice should clearly be guided by any physical
insight that we can bring to bear on the system or, more specifically,
to that particular phase of it that is under consideration. In the
absence of any other insight into the quantum many-body system it is
common to be guided by the behaviour of the corresponding classical
system (i.e., equivalently, the system when the spin quantum number $s
\rightarrow \infty$). The $XXZ$ model under consideration provides
just such an illustrative example. Thus, for $\Delta > 1$ the {\it
  classical} Hamiltonian of equation (\ref{eq:H_XXZ}) on the 2D square
lattice (and, indeed, on any bipartite lattice) is minimized by a
perfectly antiferromagnetically N\'{e}el-ordered state in the spin
$z$-direction. However, the classical gs energy is minimized by a
N\'{e}el-ordered state with spins pointing along any direction in the
spin $x$-$y$ plane (say, along the spin $x$-direction) for $-1 <
\Delta < 1$. Either of these states could be used as a CCM model state
$|\Phi\rangle$ and both are likely to be of value in different regimes
of $\Delta$ appropriate to the particular quantum phases that mimic
the corresponding classical phases. For present illustrative purposes
we restrict ourselves to the $z$-aligned N\'{e}el state as our choice
for $|\Phi\rangle$, written schematically as
\[|\Phi\rangle=|\cdots \downarrow \uparrow \downarrow \uparrow \cdots\rangle\,,\hspace{0.2in}\mbox{in the global spin axes,}\]
where $|\uparrow\rangle\equiv \bigl\vert
\frac{1}{2},+\frac{1}{2}\bigl\rangle$ and $|\downarrow\rangle\equiv
\bigl\vert \frac{1}{2},-\frac{1}{2}\bigl\rangle$ in the usual $|s,
m_{s}\rangle$ notation. Such a state is, clearly, likely to be a good
starting-point for all $\Delta > 1$, down to the expected phase
transition at $\Delta=1$ from a $z$-aligned N\'{e}el phase to an
$x$-$y$ planar N\'{e}el phase.

As indicated in section~\ref{spin_latt} it is now convenient to perform a
rotation of the axes for the up-pointing spins (i.e., those on the
sublattice with spins in the positive $z$-direction) by $180^{\circ}$
about the spin $y$-axis, so that $|\Phi\rangle$ takes the form given by
equation (\ref{local_quan}). Under this rotation, the spin operators
on the original up sub-lattice are transformed as
\begin{equation}
s^{x}\rightarrow-s^{x}, \qquad s^{y} \rightarrow s^{y}, \qquad s^{z} \rightarrow -s^{z}.
\end{equation}
The Hamiltonian of equation (\ref{eq:H_XXZ}) may thus be rewritten in
these local spin coordinate axes as
\begin{equation}
H_{XXZ} = -\frac{1}{2}\sum_{\langle i,j \rangle}[s^{+}_{i}s^{+}_{j} + s^{-}_{i}s^{-}_{j} + 2\Delta s^{z}_{i}s^{z}_{j}]\;. \label{eq:Ham_XXZ_trans}
\end{equation}

As in any application of the CCM to spin-lattice systems, we now
include in our approximations at any given order only those {\it
  fundamental configurations} that are distinct under the point and
space group symmetries of both the lattice and the Hamiltonian. The
number, $N_{f}$, of such fundamental configurations at any level of
approximation may be further restricted whenever additional
conservation laws come into play. For example, in our present case,
the $XXZ$ Hamiltonian of equation ({\ref{eq:H_XXZ}) commutes with the
  total uniform magnetization, $s^{z}_{T}=\sum^{N}_{k=1}s^{z}_{k}$, in
  the global spin coordinates, where $k$ runs over all lattice
  sites. The GS is known to lie in the $s^{z}_{T}=0$ subspace, and
  hence we exclude configurations with an odd number of spins or with
  unequal numbers of spins on the two equivalent sublattices of the
  bipartite square lattice. We show in figure~\ref{DSUB1to4_xxz}
\begin{figure}[!t]
\begin{center}                  
\includegraphics[width=10cm]{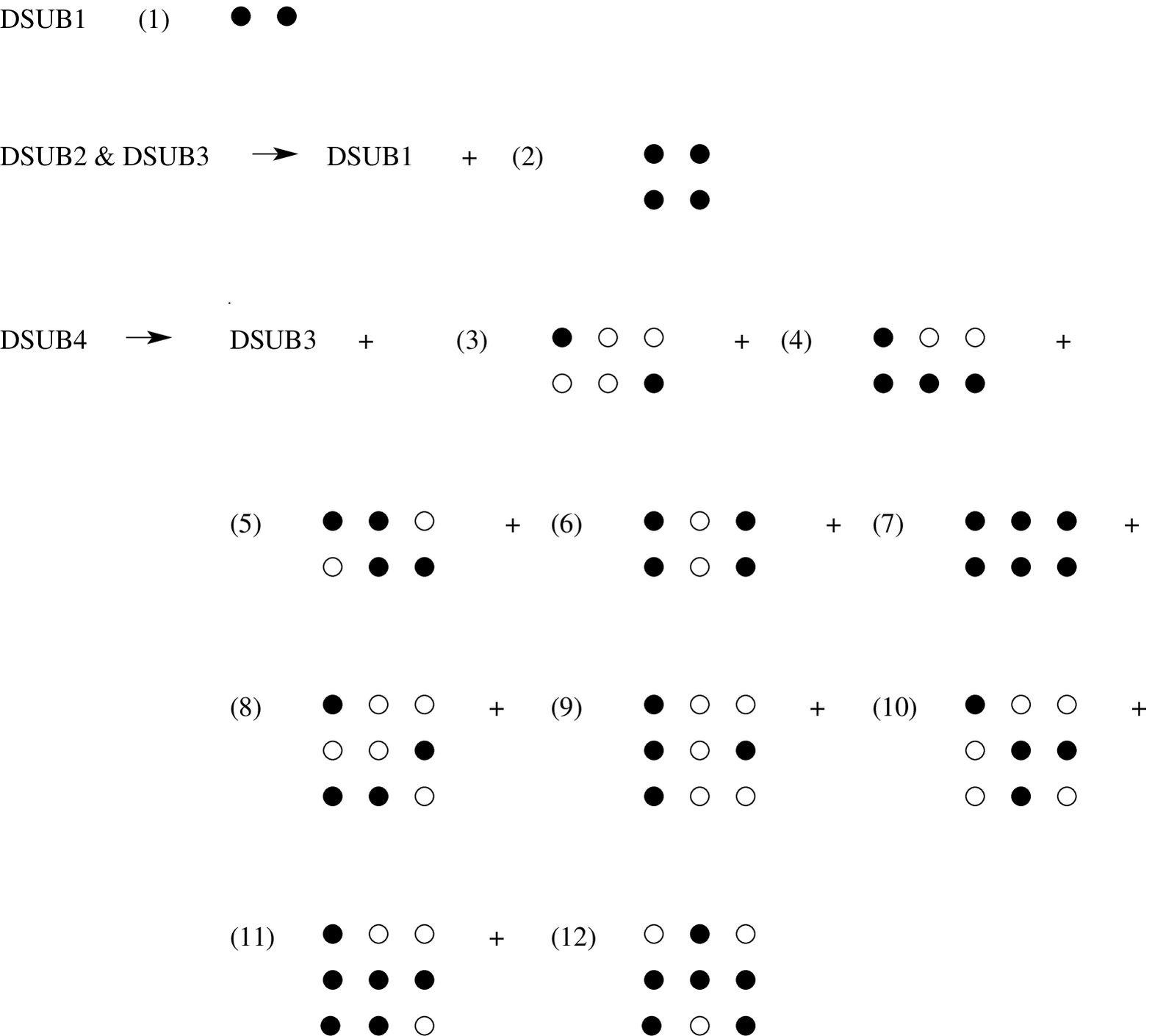} 
\vskip0.5cm
\caption{The fundamental configurations for the DSUB$m$ scheme with
  $m=\{1,2,3,4\}$ for the spin-1/2 $XXZ$ model on a square lattice in
  two dimensions. The filled circles mark the relative positions of
  the sites of the square lattice on which the spins are flipped with
  respect to the model state. The unfilled circles represent unflipped
  sites.}
\label{DSUB1to4_xxz}                                                       
\end{center}
\end{figure}   
the fundamental configurations that are accordingly allowed for the
DSUB$m$ approximations for this spin-1/2 $XXZ$ model on the 2D square
lattice, with $1 \leq m \leq 4$. We see that, for example, $N_{f}=12$
at the DSUB4 level of approximation. We also see that $N_{f}=2$ for
both the DSUB2 and DSUB3 approximations, since the conservation law
$s^{z}_{T}=0$ does not permit any additional configurations of spins
with a maximum distance $L_{3}=2$, apart from those already included
in the DSUB2 approximation.

\subsection{Ground-state energy and sublattice magnetization}
The DSUB$m$ approximations can readily be implemented for the present
spin-1/2 $XXZ$ model on the 2D square lattice for all values $m \leq
11$ with reasonably modest computing power. By comparison, the LSUB$m$
scheme can be implemented with comparable computing resources for all
values $m \leq 10$. Numerical results for the gs energy per spin and
sublattice magnetization are shown in table~\ref{Table_DSUBm_XXZ}
\begin{table}[!t]
  \caption{The ground-state energy per spin ($E/N$) and sublattice magnetization ($M$) for the spin-$1/2$ $XXZ$ model on the square lattice, obtained using the CCM DSUB$m$ approximation scheme with $1 \leq m \leq 11$ at $\Delta=1$. $N_{f}$ is the number of fundamental configurations at a given DSUB$m$ or LSUB$m$ level of approximation. $\Delta_{i} \equiv$ DSUB$m$ sublattice magnetization point of inflexion. The DSUB$m$ results for odd values of $m$, even values of $m$ and the whole series of $m$ are extrapolated separately. These results are compared to calculations using third-order spin-wave theory (SWT), exact diagonalization (ED), series expansion (SE), quantum Monte Carlo (QMC) and LSUB$\infty$ extrapolations of the CCM LSUB$m$ approximations.} 
\vskip0.2cm
\label{Table_DSUBm_XXZ}  
\begin{center}
\begin{tabular}{|c|c|c|c|c|c|c|c|} \hline\hline
{Method} & {$L_{m}$} & {$N_{f}$} & \multicolumn{2}{c|}{$E/N$} & {$M$} &  {$\Delta_{i}$} & {Max. No.} \\ [0.8ex]   \cline{4-6}   
& & & \multicolumn{3}{c|}{$\Delta = 1$} & & of spins   \\ \hline      
DSUB1=LSUB2 & 1 & 1 & \multicolumn{2}{c|}{-0.64833} & 0.421 &  0 &  2 \\ 
DSUB2 & $\sqrt{2}$ & 2 & \multicolumn{2}{c|}{-0.65311} & 0.410 & 0.258  & 4 \\  
DSUB3 & 2 & 2 & \multicolumn{2}{c|}{-0.65311} & 0.410 & 0.258 &  4 \\ 
DSUB4 & $\sqrt{5}$ & 12 & \multicolumn{2}{c|}{-0.66258} & 0.385 & 0.392 & 6 \\ 
DSUB5 & $\sqrt{8}$ & 20 & \multicolumn{2}{c|}{-0.66307} & 0.382 &  0.479 & 8  \\ 
DSUB6 & 3 & 43 & \multicolumn{2}{c|}{-0.66511} & 0.375 & 0.506 &  8   \\ 
DSUB7 & $\sqrt{10}$ & 135 & \multicolumn{2}{c|}{-0.66589} & 0.371 &  0.629  & 12  \\ 
DSUB8 & $\sqrt{13}$ & 831 & \multicolumn{2}{c|}{-0.66704} & 0.363 &  0.614  & 14  \\ 
DSUB9 & 4 & 1225 & \multicolumn{2}{c|}{-0.66701} & 0.363 & 0.654 &  14   \\ 
DSUB10 & $\sqrt{17}$ & 6874 & \multicolumn{2}{c|}{-0.66774} &  0.357 & 0.637 & 16   \\ 
DSUB11 & $\sqrt{18}$ & 14084 & \multicolumn{2}{c|}{-0.66785} & 0.356 & - $^{a}$  & 16   \\ \hline
LSUB8 &  & 1287 & \multicolumn{2}{c|}{-0.66817} & 0.352 &  & 8  \\ 
LSUB10 &  & 29605 & \multicolumn{2}{c|}{-0.66870} & 0.345 &  & 10  \\ \hline\hline   
& \multicolumn{7}{|c|}{Extrapolations}  \\ \hline      
& \multicolumn{2}{c|}{Based on} & \multicolumn{2}{c|}{$E/N$}  & \multicolumn{2}{c|}{$M$} & $\Delta_{i}$ \\  \hline          
DSUB$\infty$ & \multicolumn{2}{c|}{$m=\{6,8,10\}$} & \multicolumn{2}{c|}{-0.67082} & \multicolumn{2}{c|}{0.308} & 1.009   \\ 
DSUB$\infty$ & \multicolumn{2}{c|}{$m=\{5,7,9,11\}$} & \multicolumn{2}{c|}{-0.67122} & \multicolumn{2}{c|}{0.311} & 1.059 $^{b}$  \\ 
DSUB$\infty$ & \multicolumn{2}{c|}{$m=\{7,9,11\}$} & \multicolumn{2}{c|}{-0.66978} & \multicolumn{2}{c|}{0.319}  &\\ 
DSUB$\infty$ & \multicolumn{2}{c|}{$3\leq m \leq 11$} & \multicolumn{2}{c|}{-0.67177} & \multicolumn{2}{c|}{0.315} & 1.025 $^{c}$ \\ 
DSUB$\infty$ & \multicolumn{2}{c|}{$4 \leq m \leq 11$} & \multicolumn{2}{c|}{-0.66967} & \multicolumn{2}{c|}{0.325} & 0.979 $^{c}$ \\   \hline
LSUB$\infty$~\cite{Fa:2008} & \multicolumn{2}{c|}{$m=\{3,5,7,9\}$} & \multicolumn{2}{c|}{-0.67029} & \multicolumn{2}{c|}{0.305} &  \\ 
LSUB$\infty$~\cite{Fa:2008,Ri:2008} & \multicolumn{2}{c|}{$m=\{4,6,8,10\}$} & \multicolumn{2}{c|}{-0.66966} & \multicolumn{2}{c|}{0.310} & \\ 
LSUB$\infty$ & \multicolumn{2}{c|}{$m=\{6,8,10\}$} & \multicolumn{2}{c|}{-0.66962} & \multicolumn{2}{c|}{0.308} &  \\ \hline\hline
SWT~\cite{Ha:1992} & \multicolumn{2}{c|}{} & \multicolumn{2}{c|}{-0.66999} & \multicolumn{2}{c|}{0.307} &  \\  
SE~\cite{Rit:2004} & \multicolumn{2}{c|}{} & \multicolumn {2}{c|}{-0.6693} & \multicolumn{2}{c|}{0.307} &  \\  
ED~\cite{Zh:1991} & \multicolumn{2}{c|}{} & \multicolumn{2}{c|}{-0.6700} & \multicolumn{2}{c|}{0.3173} &  \\ 
QMC~\cite{Sa:1997} & \multicolumn{2}{c|}{} & \multicolumn{2}{c|}{-0.669437(5)} & \multicolumn{2}{c|}{0.3070(3)} &  \\   \hline\hline 
\end{tabular}
\end{center}
\protect \underline{NOTES}: \\  
\protect  $^{a}$ The magnetization point of inflexion for DSUB11 is not available since we only calculated at $\Delta=1$ in this approximation.  \\ 
\protect $^{b}$ The magnetization points of inflexion for the odd DSUB$m$ levels are extrapolated using $m=\{5,7,9\}$.  \\ 
\protect $^{c}$ The magnetization points of inflexion for the whole series of DSUB$m$ data are extrapolated as indicated, but without $m=11$.    
\end{table} 
at the isotropic point $\Delta=1$ at various levels of approximation,
and corresponding results for the same quantities are displayed
graphically in figures \ref{DSUBm_XXZ_E} and \ref{DSUBm_XXZ_M} as functions of the anisotropy parameter $\Delta$.
\begin{figure}[!b]  
\begin{center}
\includegraphics[width=8cm]{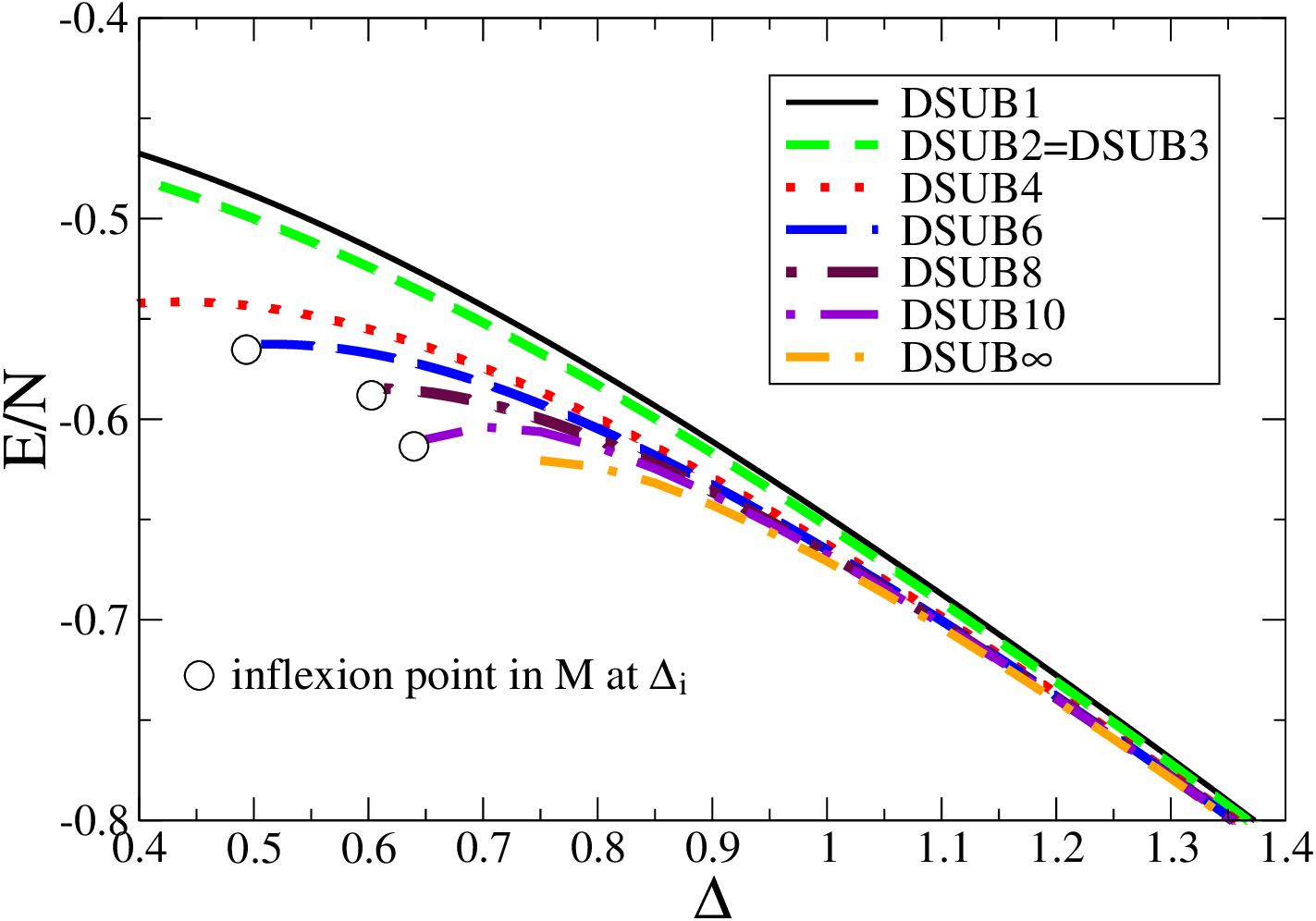} 
\caption{(Color online) CCM results for the ground-state energy per
  spin, $E/N$, as a function of the anisotropy parameter $\Delta$, of
  the spin-$1/2$ $XXZ$ model on the square lattice, using various
  DSUB{\it m} approximations based on the {\it z}-aligned N\'{e}el
  model state. The DSUB$m$ results with $m=\{6,8,10\}$ are
  extrapolated using the leading (linear) fit of equation
  (\ref{Extrapo_E}) and shown as the curve
  DSUB$\infty$. $\Delta_{i}\equiv$ magnetization point of inflexion,
  described in the text.}
\vskip0.1cm
\label{DSUBm_XXZ_E}
\end{center}
\end{figure}
\begin{figure}[!t] 
\begin{center}
\vskip0.5cm
\includegraphics[width=8cm]{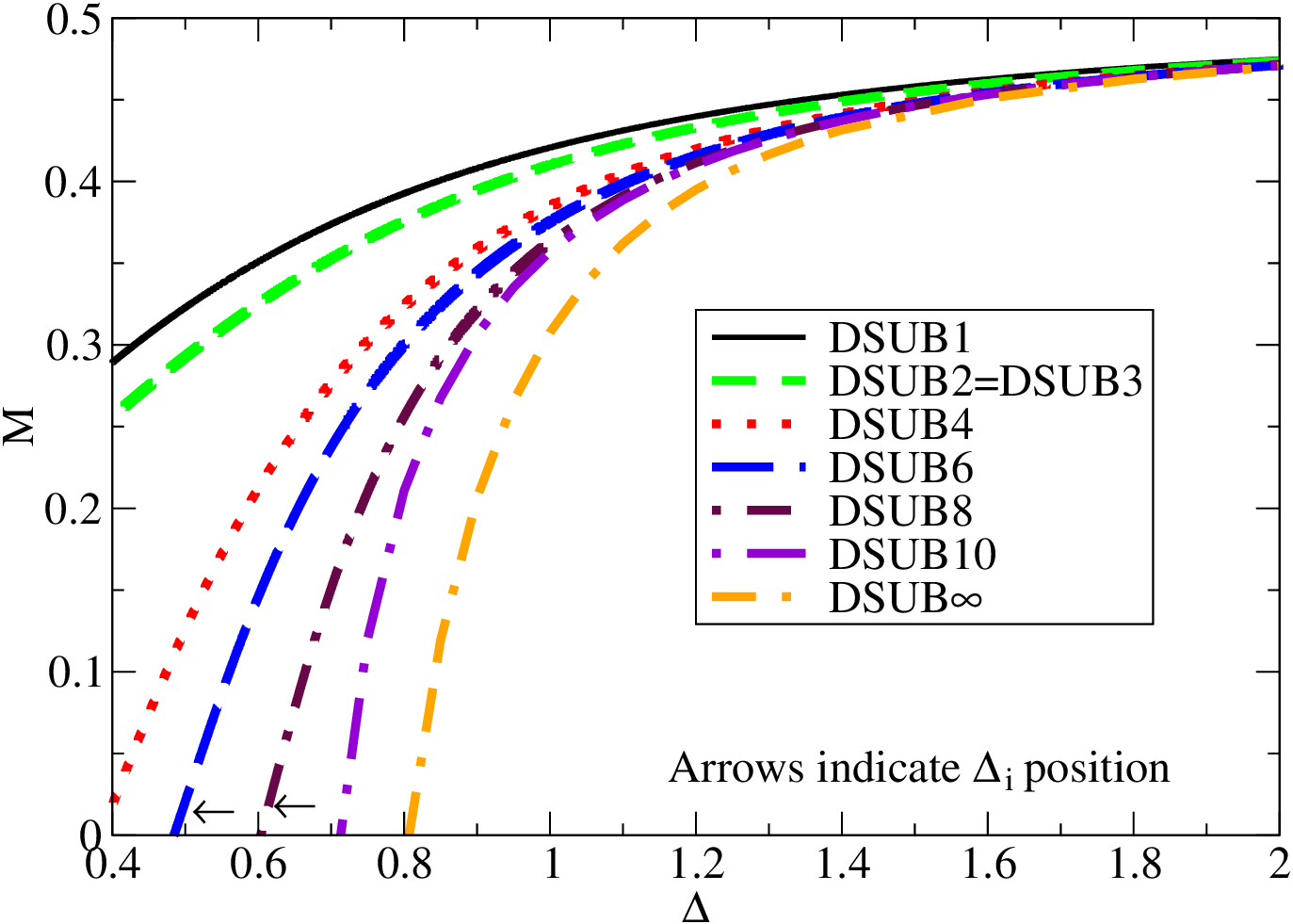}  
\caption{(Color online) CCM results for the sublattice
  magnetization, $M$, as a function of the anisotropy parmeter
  $\Delta$, of the spin-$1/2$ $XXZ$ model on the square lattice, using
  various DSUB{\it m} approximations based on the {\it z}-aligned N\'{e}el
  model state. The DSUB$m$ results with $m=\{6,8,10\}$ are
  extrapolated using the leading (linear) fit of equation
  (\ref{Extrapo_M}) and shown as the curve
  DSUB$\infty$. $\Delta_{i}\equiv$ point of inflexion in the curve,
  shown by arrows in the figure.}
\label{DSUBm_XXZ_M}
\end{center}
\end{figure}    

We also show in table~\ref{Table_DSUBm_XXZ} for the isotropic
Heisenberg Hamiltonian ($\Delta=1$) the results for the gs energy and
sublattice magnetization using the leading (linear) extrapolation
schemes of equations (\ref{Extrapo_E}) and (\ref{Extrapo_M})
respectively of the DSUB$m$ data, employing various subsets of
results. Comparison is also made with corresponding LSUB$m$
extrapolation schemes for the same model~\cite{Fa:2008,Ri:2008}. The
results are generally observed to be in excellent agreement with each
other, even though the DSUB$\infty$ extrapolations have employed the
simple leading (linear) fits of equations (\ref{Extrapo_E}) and
(\ref{Extrapo_M}), whereas the corresponding LSUB$\infty$ results
shown~\cite{Fa:2008,Ri:2008} have been obtained from the potentially
more accurate quadratic fits
$E/N=a_{0}+a_{1}(1/m^{2})+a_{2}(1/m^{2})^{2}$,
$M=b_{0}+b_{1}(1/m^{2})+b_{2}(1/m)^{2}$, to the LSUB$m$ data, in which
the next-order (quadratic) corrections have also been included in the
relevant expansion parameters, $1/m^{2}$ and $1/m$,
respectively. Excellent agreement of all the CCM extrapolations is
also obtained with the results from the best of the alternative
methods for this model, including third-order spin-wave theory
(SWT)~\cite{Ha:1992}, linked-cluster series expansion
techniques~\cite{Rit:2004}, and the extrapolations to infinite lattice
size ($N \rightarrow \infty$) from the exact diagonalization (ED) of
small lattices~\cite{Zh:1991}, and quantum Monte Carlo (QMC)
calculations for larger lattices~\cite{Mo:1953}.

We note that it has been observed and well documented in the past (and
see, e.g., Ref.~\cite{Fa:2008}) that the CCM LSUB$m$ results for this
model (and many others) for both the gs energy $E$ and the sublattice
magnetization $M$ show a distinct period-2 ``staggering'' effect with
index $m$, according to whether $m$ is even or odd. As a consequency
the LSUB$m$ data for both $E$ and $M$ converge differently for the
even-$m$ and the odd-$m$ sequences, similar to what is observed very
frequently in perturbation theory in corresponding even and odd
orders~\cite{Mo:1953}. As a rule, therefore, the LSUB$m$ data are
generally extrapolated separately for even $m$ and for odd values of
$m$, since the staggering makes extrapolations using both odd and even
values together extremely difficult. We show in
figure~\ref{DSUBm_staggered_all}
\begin{figure}[!t]
\begin{center}
\mbox{
  \subfigure[Ground-state energy per spin]{\label{DSUBm_E_staggered_XXZ}\includegraphics[scale=0.3,angle=270]{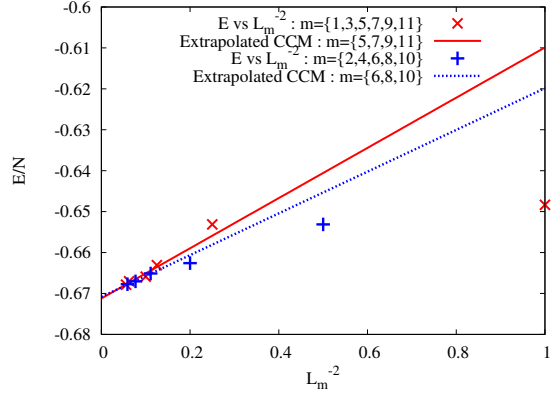}} 
}
\mbox{
  \subfigure[Ground-state sublattice magnetization]{\label{DSUBm_M_staggered_XXZ}\includegraphics[scale=0.3,angle=270]{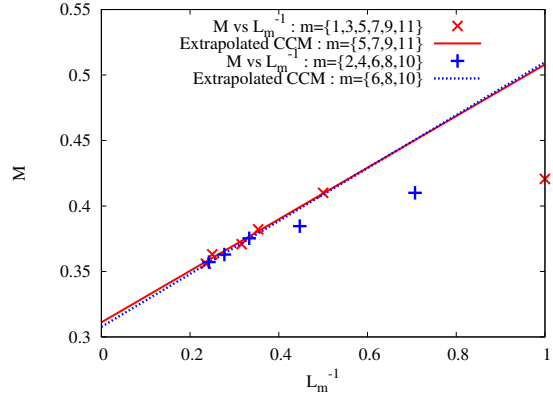}} 
}
\caption{(Color online) Illustration of the staggered nature of the
  DSUB$m$ scheme for the gs energy per spin, $E/N$, and sublattice magnetization, $M$, for the
  spin-$1/2$ antiferromagnetic $XXZ$ model on the square lattice. The DSUB$m$ data are plotted against $1/L_{m}^{2}$ for $E/N$ and against $1/L_{m}$ for $M$. The results clearly justify the heuristic extrapolation schemes of equations (\ref{Extrapo_E}) and (\ref{Extrapo_M}).}
\label{DSUBm_staggered_all}
\end{center}
\end{figure}         
our DSUB$m$ results for the gs energy per spin and the sublattice
magnetization plotted against $1/L^{2}_{m}$ and $1/L_{m}$,
respectively.  The higher $m$ values clearly cluster well in both
cases on straight lines, thereby justifying {\it a posteriori} our
heuristic extrapolation fits of equations (\ref{Extrapo_E}) and
(\ref{Extrapo_M}). Just as in the LSUB$m$ case a slight ``even-odd
staggering'' effect is observed in the DSUB$m$ data (perhaps more so
for the energy than for the sublattice magnetization), although it is
less pronounced than for the corresponding LSUB$m$
data~\cite{Fa:2008}.

\subsection{Termination or critical points}
\label{DSUBm_XXZ_termination}
Before discussing our DSUB$m$ results further for this model we note
that the comparable LSUB$m$ solutions actually terminate at a critical
value $\Delta_{c}=\Delta_{c}(m)$, which itself depends on the truncation index
$m$~\cite{Fa:2004}. Such LSUB$m$ termination points are very common
for many spin-lattice systems and are very well documented and
understood (and see, e.g., Ref.~\cite{Fa:2004}). In all such cases a
termination point always arises due to the solution of the CCM
equations becoming complex at this point, beyond which there exist two
branches of entirely unphysical complex conjugate
solutions~\cite{Fa:2004}. In the region where the solution reflecting
the true physical solution is real there actually also exists another
(unstable) real solution. However, only the (shown) upper branch of
these two solutions reflects the true (stable) physical GS, whereas
the lower branch does not.  The physical branch is usually easily
identified in practice as the one which becomes exact in some known
(e.g., perturbative) limit.  This physical branch then meets the
corresponding unphysical branch at some termination point beyond which
no real solutions exist. The LSUB$m$ termination points are themselves
also reflections of the quantum phase transitions in the real system,
and may be used to estimate the position of the phase
boundary~\cite{Fa:2004}.

It is interesting and intriguing to note that when the DSUB$m$
approximations are applied to the $XXZ$ model, they do not terminate
as do the corresponding LSUB$m$ approximations. We have no real
explanation for this rather striking difference in behaviour for two
apparently similar schemes applied to the same model. However, it is
still possible to use our DSUB$m$ data to extract an estimate for the
physical phase transition point at which the $z$-aligned N\'{e}el
phase terminates. As has been justified and utilized
elsewhere~\cite{Fa:1997}, a point of inflexion at $\Delta=\Delta_{i}$
in the sublattice magnetization $M$ as a function of $\Delta$ clearly
indicates the onset of an instability in the system. Such inflexion
points $\Delta_{i}=\Delta_{i}(m)$ occur for all DSUB$m$ approximations, as indicated
in table~\ref{Table_DSUBm_XXZ} and figure~\ref{DSUBm_XXZ_M}. The
DSUB$m$ approximations are thus expected to be unphysical for $\Delta
< \Delta_{i}(m)$, and we hence show the corresponding results for the
gs energy per spin in figure~\ref{DSUBm_XXZ_E} only for values
$\Delta_{i} > \Delta_{i}(m)$. Heuristically, we find that the
magnetization inflexion points $\Delta_{i}(m)$ scale linearly with
$1/L_{m}$ to leading order, and the extrapolated results shown in
table~\ref{Table_DSUBm_XXZ} have been performed with the leading
(linear) fit, $\Delta_{i}=c_{o}+c_{1}(1/L_{m})$, commensurate with the
corresponding linear fits in $(1/L^{2}_{m})$ and $(1/L_{m})$ for the
gs energy per spin and sublattice magnetization of equations
(\ref{Extrapo_E}) and (\ref{Extrapo_M}), respectively. All of the
various extrapolations shown in table~\ref{Table_DSUBm_XXZ} for
$\Delta_{i}$ in the limit $m \rightarrow \infty$ are in good agreement
with one another, thereby again demonstrating the robust quality of
the heuristic extrapolation scheme. Furthermore, they are also in
excellent agreement with the expected phase transition point at
$\Delta_{c} \equiv 1$ between two quasiclassical N\'{e}el-ordered
phases aligned along the spin $z$-axis (for $\Delta > 1$) and in some
arbitrary direction in the spin $x$-$y$-plane (for $|\Delta| < 1$).

Although we do not do so here, the $x$-$y$ planar N\'{e}el phase could
itself also easily be investigated by another CCM DSUB$m$ series of
calculations based on a model state $|\Phi\rangle$ with perfect
N\'{e}el ordering in, say, the $x$-direction.

Summarizing our results so far, we observe that the DSUB$m$ scheme has
at, least partially, fulfilled the expectations placed upon it for the
present model. Accordingly, we now apply it to the second test model
of the spin-1/2 $XY$ model on the 2D square lattice.

\section{The spin-1/2 $XY$ model on the square lattice}
\label{XY}
The Hamiltonian of the $XY$ model~\cite{Fa:1997} in global spin
coordinates, is written as
\begin{equation}
H_{XY} = \sum_{\langle i,j \rangle}[(1+\Delta)s^{x}_{i}s^{x}_{j} + (1-\Delta)s^{y}_{i}s^{y}_{j}]\,;\hspace{0.2in} -1 \leq \Delta \leq 1\,,  \label{eq:H_XY}
\end{equation}
where the sum on $\langle i,j \rangle$ again runs over all
nearest-neighbour pairs of lattice sites and counts each pair only
once. We again consider the case of spin-1/2 particles on each site of
an infinite square lattice.

For the classical model described by equation (\ref{eq:H_XY}), it is
clear that the GS is a N\'{e}el state in the $x$-direction for $0 <
\Delta \leq 1$ and a N\'{e}el state in the $y$-direction for $-1 \leq
\Delta < 0$. Hence, since we only consider the case $0 \leq \Delta
\leq 1$, we choose as our CCM model state $|\Phi\rangle$ for the
quantum $XY$ model a N\'{e}el state aligned along the $x$-direction,
written schematically as,
\[|\Phi\rangle=|\cdots \leftarrow \, \rightarrow \, \leftarrow \, \rightarrow
\cdots\rangle\;,\hspace{0.2in}\mbox{in the global spin axes}.\] 
Clearly the case $-1 \leq \Delta < 0$ is readily obtained from the case $0
< \Delta \leq 1$ by interchange of the $x$- and $y$-axes.

Once again we now perform our usual rotation of the spin axes on each
lattice site so that $|\Phi\rangle$ takes the form given by equation
(\ref{local_quan}}) in the rotated local spin coordinate frame. Thus,
for the spins on the sublattice where they point in the negative
$x$-direction in the global spin axes (i.e., the left-pointing spins)
we perform a rotation of the spin axes by $+90^{\circ}$ about the spin
$y$-axis. Similarly, for the spins on the other sublattice where they
point in the positive $x$-direction in the global spin axes (i.e., the
right-pointing spins) we perform a rotation of the spin axes by
$-90^{\circ}$ about the spin $y$-axis. Under these rotations the spin
operators are transformed as
\begin{subequations} 
\begin{eqnarray}    
s^{x} \rightarrow  s^{z}\;, \qquad s^{y}  \rightarrow s^{y}\;,  \qquad s^{z} \rightarrow -s^{x}\;, \qquad \mbox{left-pointing spins}; \\
s^{x} \rightarrow -s^{z}\;, \qquad s^{y}  \rightarrow s^{y}\;, \qquad  s^{z} \rightarrow s^{x}\;, \qquad  \mbox{right-pointing spins}.
\end{eqnarray} 
\end{subequations}

The Hamiltonian of equation (\ref{eq:H_XY}) may thus be rewritten in the local spin coordinate axes defined above as
\begin{equation}
H_{XY} = \sum_{\langle i,j \rangle}\left[-(1+\Delta)s^{z}_{i}s^{z}_{j}-\frac{1}{4}(1-\Delta)(s^{+}_{i}s^{+}_{j}+s^{-}_{i}s^{-}_{j})+\frac{1}{4}(1-\Delta)(s^{+}_{i}s^{-}_{j}+s^{-}_{i}s^{+})\right] \;. \label{Hamilton_XY_trans}
\end{equation}

\begin{figure}[!t]    
\begin{center}
\includegraphics[width=10cm]{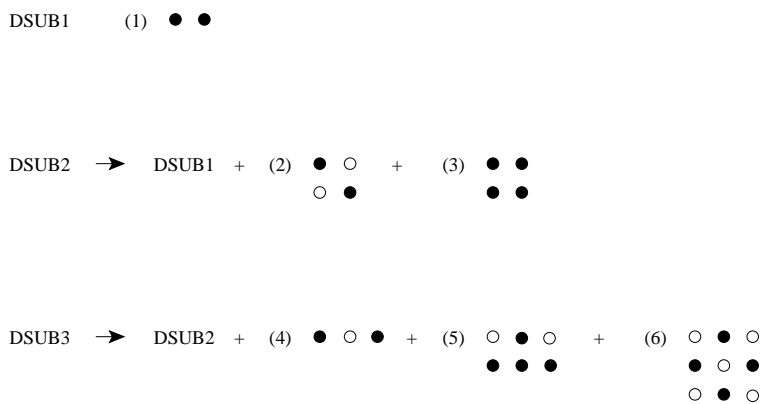} 
\vskip0.5cm
\caption{The fundamental configurations for the DSUB$m$ scheme with
  $m=\{1,2,3\}$ for the spin-1/2 $XY$ model on a square lattice in two dimensions. The filled circles mark the
  relative positions of the sites of the square lattice on which the
  spins are flipped with respect to the model state.  The unfilled
  circles represent unflipped sites.}
\label{DSUB1to3_xy}
\end{center}
\end{figure}

As before, we now have to evaluate the fundamental configurations that
are retained in the CCM correlation operators $S$ and $\tilde{S}$ at
each DSUB$m$ level of approximation. Although the point and space
group symmeries of the square lattice (common to both the $XXZ$ and
$XY$ models considered here) and the two Hamiltonians of equations
(\ref{eq:Ham_XXZ_trans}) and (\ref{Hamilton_XY_trans}) are identical,
the numbers $N_{f}$ of fundamental configurations for a given DSUB$m$
level are now larger (except for the case $m=1$) for the $XY$ model
than for the $XXZ$ model, since the uniform magnetization is no longer
a good quantum number for the $XY$ model, $[H_{XY},S^{z}_{T}]\neq
0$. Nevertheless, we note from the form of equation
(\ref{Hamilton_XY_trans}), in which the spin-raising and spin-lowering
operators appear only in combinations that either raise or lower the
number of spin flips by two (viz., the $s^{+}_{i}s^{+}_{j}$ and
$s^{-}_{i}s^{-}_{j}$ combinations, respectively) or leave them
unchanged (viz., the $s^{+}_{i}s^{-}_{j}$ and $s^{-}_{i}s^{+}_{j}$
combinations), it is only necessary for the $s^{z}_{T}=0$ GS to
consider fundamental configurations that contain an even number of
spins. Thus, the main difference for the $XY$ model over the $XXZ$
model is that we must now also consider fundamental configurations in
which we drop the restriction for the former case of having an equal
number of spins on the two equivalent sublattices of the bipartite
square lattice that was appropriate for the latter case. We show in
figure~\ref{DSUB1to3_xy} the fundamental configurations that are
allowed for the spin-1/2 $XY$ model on the square lattice for the
DSUB$m$ approximation with $1 \leq m \leq 3$, and we invite the reader
to compare with the corresponding fundamental configurations for the
spin-1/2 $XXZ$ model on the same square lattice shown in figure 1.
The corresponding numbers $N_{f}$ of fundamental configurations for
the $XY$ model are also shown in table~\ref{table_DSUBm_XY} for the
higher DSUB$m$ approximations with $m \leq 9$ for which we present
results below.

\subsection{Ground-state energy and sublattice magnetization}
\begin{figure}[!t]
\begin{center}
\vskip0.2cm
\includegraphics[width=8cm]{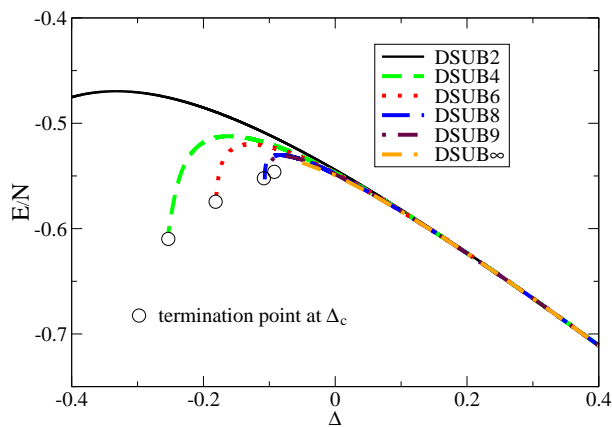}
\caption{(Color online) CCM results for the gs energy of the
  spin-$1/2$ $XY$ model on the square lattice obtained using the
  DSUB{\it m} approximation based on the N\'{e}el state aligned along
  any axis in the $x$-$y$ plane.  The DSUB$m$ results with $m=\{5,7,9\}$
  are extrapolated using equation (\ref{Extrapo_E}) to give the curve labelled DSUB$\infty$.}
\vskip0.2cm
\label{DSUBm_XY_E}
\end{center}
\end{figure}

\begin{figure}[!t]
\begin{center}
\includegraphics[width=8cm]{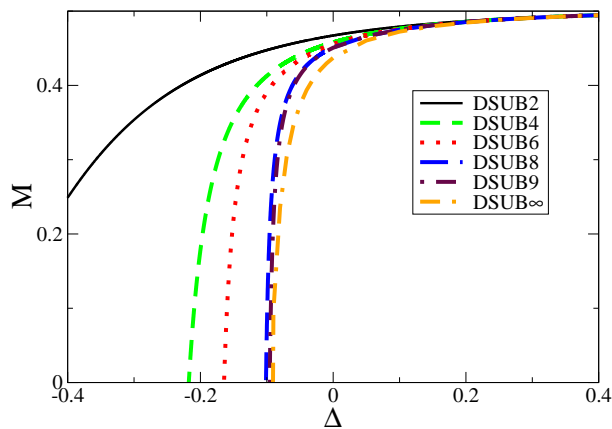}
\caption{(Color online) CCM results for the gs sublattice
  magnetization of the spin-$1/2$ $XY$ model on the square lattice
  obtained using various DSUB{\it m} approximations based on the
  N\'{e}el state aligned along any axis in the $x$-$y$ plane.  The
  DSUB$m$ results with $m=\{5,7,9\}$ are extrapolated using equation
  (\ref{Extrapo_M}) to give the curve labelled DSUB$\infty$.}
\label{DSUBm_XY_M}
\end{center}
\end{figure}

We present results for the spin-1/2 $XY$ model on the square lattice
in the CCM DSUB$m$ approximations for all values $m \leq 9$ that can
be easily computed with very modest computing power. Comparable
computing power enables the corresponding LSUB$m$ scheme to be
implemented for all $m \leq 8$. Numerical results for the gs energy
per spin and sublattice magnetization are shown in
table~\ref{table_DSUBm_XY} at the isotropic point at $\Delta=0$ at
various levels of approximation, and corresponding results for the
same gs quantities are shown graphically in figures~\ref{DSUBm_XY_E}
and~\ref{DSUBm_XY_M} as functions of the anisotropy parameter
$\Delta$.
\begin{table}[!t]           
  \caption{The ground-state energy per spin ($E/N$) and sublattice magnetization ($M$) for the spin-$1/2$ $XY$ model on the square lattice, obtained using the CCM DSUB$m$ approximation scheme with $1 \leq m \leq 9$ at $\Delta=0$. $N _{f}$ is the number of fundamental configurations at a given level of DSUB$m$ or LSUB$m$ approximation. $\Delta_{c} \equiv$ DSUB$m$ termination point. The DSUB$m$ results for odd values of $m$, even values of $m$ and the whole series of $m$ are extrapolated separately. These results are compared to calculations of series expansion (SE), quantum Monte Carlo (QMC) and LSUB$\infty$ extrapolations of the CCM LSUB$m$ approximations.} 
\label{table_DSUBm_XY} 
\vskip0.2cm
\begin{center}
\begin{tabular}{|c|c|c|c|c|c|c|c|} \hline\hline
Method & $L_{m}$ & $N_{f}$ & \multicolumn{2}{c|}{$E/N$} & $M$ & $\Delta_{c}$ & Max. No. \\ \cline{4-6}
 & &  & \multicolumn{3}{c|}{$\Delta = 0$} & & of spins  \\ \hline
DSUB1=LSUB2 & 1 & 1  & \multicolumn{2}{c|}{-0.54031} & 0.475 &  & 2 \\   
DSUB2 & $\sqrt{2}$  & 3 & \multicolumn{2}{c|}{-0.54425} &  0.467 &   & 4 \\  
DSUB3 & 2 & 6 & \multicolumn{2}{c|}{-0.54544} &  0.464  &  & 4 \\ 
DSUB4 & $\sqrt{5}$ & 21 & \multicolumn{2}{c|}{-0.54724} &  0.458 &  -0.253 & 6 \\ 
DSUB5 & $\sqrt{8}$ & 44 & \multicolumn{2}{c|}{-0.54747} & 0.456 & -0.205 & 8\\ 
DSUB6 & 3 & 78 & \multicolumn{2}{c|}{-0.54774} & 0.455  & -0.181 & 8 \\ 
DSUB7 & $\sqrt{10}$ & 388 & \multicolumn{2}{c|}{-0.54811} & 0.453 & -0.135 & 12 \\ 
DSUB8 & $\sqrt{13}$ & 1948 & \multicolumn{2}{c|}{-0.54829} & 0.451 & -0.107 & 14 \\ 
DSUB9 & 4 &  3315 & \multicolumn{2}{c|}{-0.54833} & 0.451 & -0.099 & 14 \\ \hline   
LSUB6 &  & 131 & \multicolumn{2}{c|}{-0.54833} & 0.451 & -0.073 & 6 \\    
LSUB8 &  & 2793 & \multicolumn{2}{c|}{-0.54862} & 0.447 & -0.04 & 8 \\ \hline\hline
& \multicolumn{7}{|c|}{Extrapolations}  \\ \hline
& \multicolumn{2}{c|}{Based on} & \multicolumn{2}{c|}{$E/N$} & \multicolumn{2}{c|}{$M$} & $\Delta_{c}$ \\ \hline           
DSUB$\infty$ & \multicolumn{2}{c|}{$m=\{4,6,8\}$} & \multicolumn{2}{c|}{-0.54879} & \multicolumn{2}{c|}{0.442} & -0.036   \\ 
DSUB$\infty$ & \multicolumn{2}{c|}{$m=\{5,7,9\}$} & \multicolumn{2}{c|}{-0.54923} & \multicolumn{2}{c|}{0.437} & 0.011 \\  
DSUB$\infty$ & \multicolumn{2}{c|}{$4 \leq m \leq 9$} & \multicolumn{2}{c|}{-0.54884} & \multicolumn{2}{c|}{0.441} & -0.029  \\  \hline
LSUB$\infty$~\cite{Fa:1997} & \multicolumn{2}{c|}{$m=\{4,6,8\}$}  & \multicolumn{2}{c|}{-0.54892} & \multicolumn{2}{c|}{0.435} & 0.00 \\ \hline\hline       
SE~\cite{Ha:1991} & \multicolumn{2}{c|}{} & \multicolumn{2}{c|}{-0.54883} & \multicolumn{2}{c|}{0.43548} & 0.0 \\    
QMC~\cite{Sa:1999} & \multicolumn{2}{c|}{} & \multicolumn{2}{c|}{-0.548824(2)} & \multicolumn{2}{c|}{0.437(2)} &  \\ \hline\hline
\end{tabular}
\end{center}   
\vskip0.5cm
\end{table}

We also show in table~\ref{table_DSUBm_XY} for the isotropic $XY$
Hamiltonian ($\Delta=0$) the results for the gs energy and sublattice
magnetization using the leading (linear) extrapolation schemes of
equations (\ref{Extrapo_E}) and (\ref{Extrapo_M}) respectively of the
DSUB$m$ data, employing various subsets of our results, as for the
$XXZ$ model considered previously. We also compare in
table~\ref{table_DSUBm_XY} the present results with the corresponding
CCM LSUB$m$ results~\cite{Fa:1997} for the same model. All of the CCM
results are clearly in excellent agreement both with one
another and with the results of best of the alternative methods
available for this model, including the linked-cluster series
expansion (SE) techniques~\cite{Ha:1991} and a quantum Monte Carlo
(QMC) method~\cite{Sa:1999}.

\begin{figure}[!t]
\begin{center}            
\mbox{
	\subfigure[Ground-state energy per spin]{\label{DSUBm_E_staggered_XY}\includegraphics[scale=0.30,angle=270]{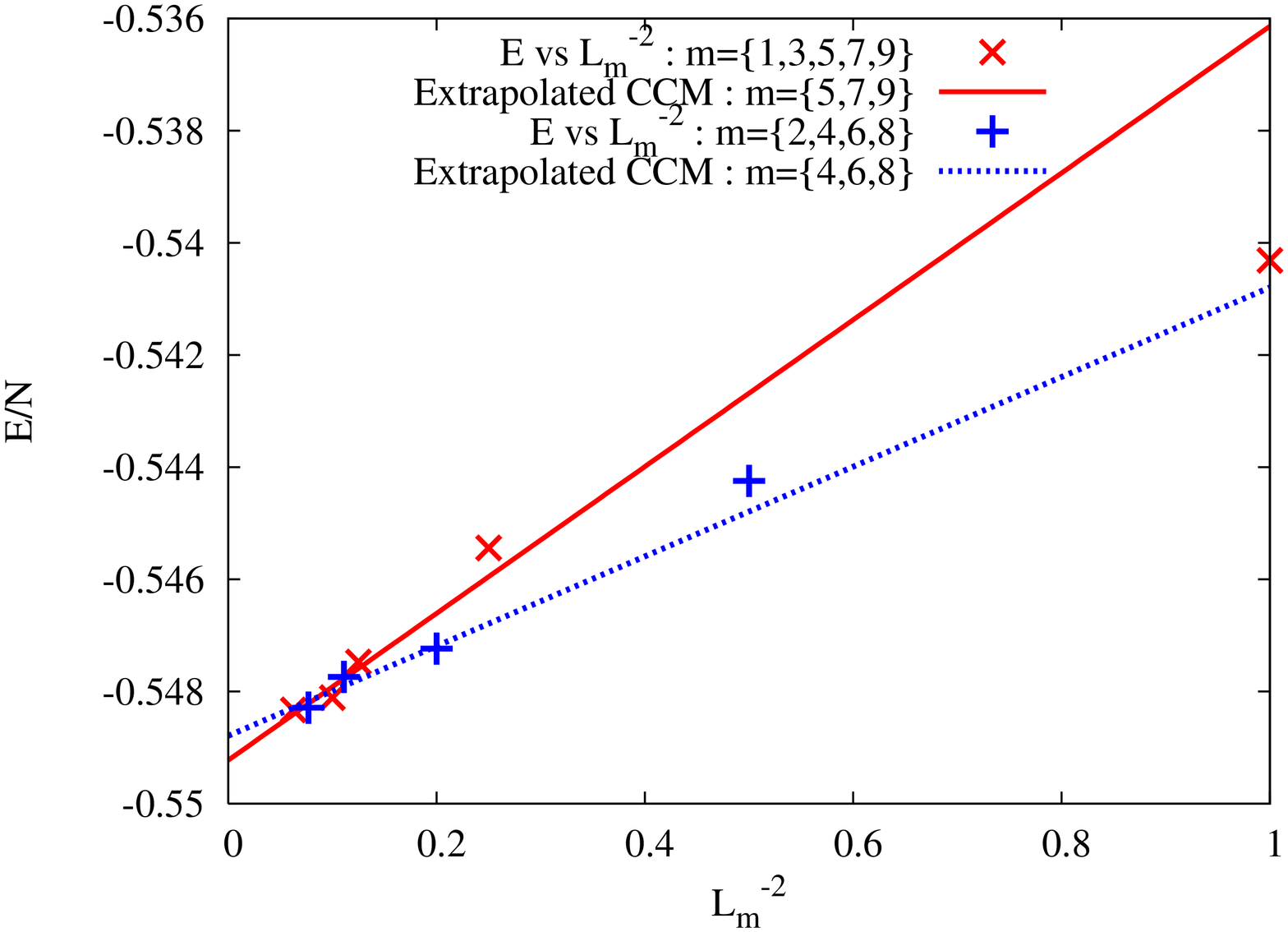}} 
}
\mbox{   
	\subfigure[Ground-state sublattice magnetization]{\label{DSUBm_M_staggered_XY}\includegraphics[scale=0.30,angle=270]{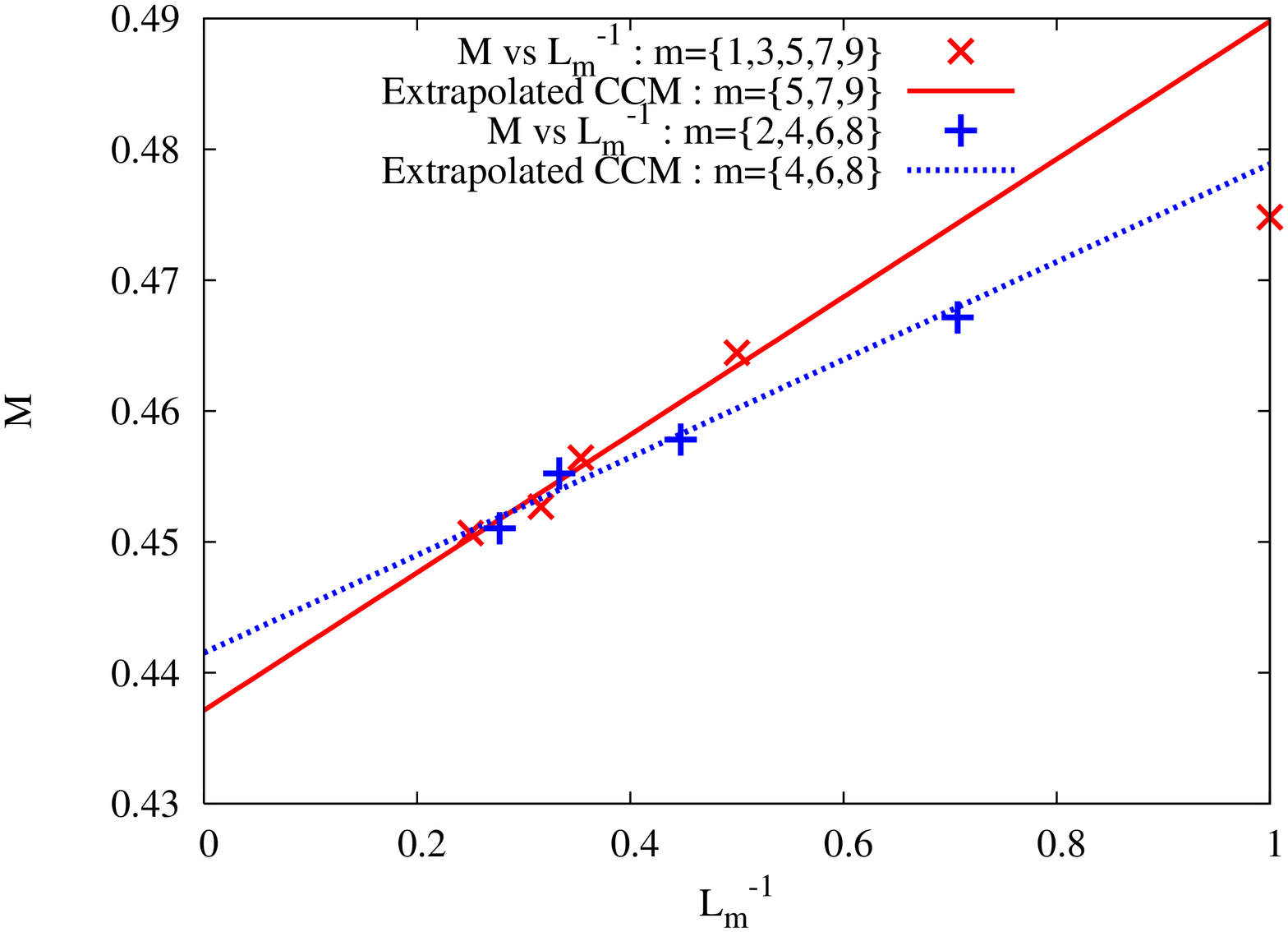}} 
}
\end{center}
\caption{(Color online) Illustration of the staggered nature of the
  DSUB$m$ scheme for the gs energy per spin, $E/N$, and sublattice
  magnetization, $M$, for the spin-$1/2$ $XY$ model on the square
  lattice. The DSUB$m$ data are plotted against $1/L_{m}^{2}$ for
  $E/N$ and against $1/L_{m}$ for $M$. The results clearly justify the
  heuristic extrapolation schemes of equations (\ref{Extrapo_E}) and
  (\ref{Extrapo_M}).}
\label{DSUBm_staggered_XY_all}
\end{figure}

We again show in figure~\ref{DSUBm_staggered_XY_all} our DSUB$m$
results for the present $XY$ model for the gs energy per spin and the
sublattice magnetization, plotted respectively against $1/L^{2}_{m}$
and $1/L_{m}$.
As previously for the $XXZ$ model, the higher $m$ values cluster well
on straight lines in both cases, thereby justifying once more our
heursitic choice of extrapolation fits indicated in equations
(\ref{Extrapo_E}) and
(\ref{Extrapo_M}). Figures~\ref{DSUBm_E_staggered_XY} and
\ref{DSUBm_M_staggered_XY} again show an ``even-odd'' staggering
effect in the termination index $m$ for the DSUB$m$ data, which is
perhaps slightly more pronounced than that for the $XXZ$ model shown
in figures~\ref{DSUBm_E_staggered_XXZ} and
\ref{DSUBm_M_staggered_XXZ}. For this reason we have again shown
separate extrapolations of our DSUB$m$ results in
table~\ref{table_DSUBm_XY} for the even-$m$ data and the odd-$m$ data,
as well as results using all (higher) values of $m$.

\subsection{Termination or critical points}
It is interesting to note that for the present $XY$ model the CCM
DSUB$m$ solutions (with our choice of model state as a N\'{e}el state
in the $x$-direction) now do physically terminate for all values of
the truncation index $m \geq 4$ at a critical value
$\Delta_{c}=\Delta_{c}(m)$, exactly as commonly occurs (as for the
present model) for the LSUB$m$ calculations, as we explained above in
section~\ref{DSUBm_XXZ_termination}. Why such DSUB$m$
terminations occur for the $XY$ model but not for the previous $XXZ$
model is not obvious to us. The corresponding termination points,
$\Delta_{c}=\Delta_{c}(m)$, at various DSUB$m$ and LSUB$m$ levels of
approximation are shown in table~\ref{table_DSUBm_XY}. It has been
shown previously~\cite{Bi:1994} that $\Delta_{c}(m)$ scales well with
$1/m^{2}$ for the LSUB$m$ data, and the LSUB$\infty$
result~\cite{Fa:1997} shown in table~\ref{table_DSUBm_XY} was obtained
by a leading (linear) fit, $\Delta_{c}(m)=d_{0}+d_{1}(1/m^{2})$. We
find heuristically that the best large-$m$ asymptotic behaviour of the
DSUB$m$ data for $\Delta_{c}(m)$ is against $1/L^{2}_{m}$ as the
scaling parameter. Accordingly, the DSUB$\infty$ values for
$\Delta_{c}$ in table~\ref{table_DSUBm_XY} are obtained with the
leading (linear) fit, $\Delta_{c}(m)=d_{0}+d_{1}(1/L_{m}^{2})$. We see
that both the LSUB$\infty$ and DSUB$\infty$ results for $\Delta_{c}
\equiv \Delta_{c}(\infty)$ agree very well with the value
$\Delta_{c}=0$ that is known to be the correct value for the phase
transition in the one-dimensional spin-1/2 $XY$ chain from the known
exact solution~\cite{Li:1961}, and which is believed also to be the
phase transition point for higher dimensions, including the present 2D
square lattice, on symmetry grounds.

\section{Conclusions}
\label{discussion}
From the two nontrivial benchmark spin-lattice problems that we have
investigated here, it is clear that the new DSUB$m$ approximation
scheme works well for calculating their gs properties and phase
boundaries. We have utilized here only the simplest leading-order
extrapolation schemes in the pertinent scaling variables, and have
shown that these may be chosen, for example, as $1/L^{2}_{m}$ for the
gs energy and $1/L_{m}$ for the order parameter. Clearly, in general,
the results can be further improved by keeping higher-order terms in
these asymptotic expansions (i.e., by retaining higher powers in the
polynomial scaling expansions) although more data points may then be
needed, especially in cases where the ``even-odd'' staggering effect
is pronounced, as for $XY$ model presented here. For further use of
the scheme for more complex lattice models (e.g., those exhibiting
geometric or dynamic frustration) it will be necessary to re-visit the
validity of these expansions, but a great deal of previous experience
in such cases for the LSUB$m$ scheme will provide good guidance.

On the basis of the test results presented here, the DSUB$m$ scheme
clearly fulfills the first of our two main criteria for introducing
it, viz., that the number of fundamental configurations, $N_{f}$,
increases less rapidly with truncation index $m$ than for the
corresponding LSUB$m$ series of approximations. At the same time our
second criterion of capturing the physically most important
configurations at relatively low levels of approximation also seems to
be fulfilled, according to our experience with the convergence of the
DSUB$m$ sequences for observable quantities. At the very least we now
have two schemes (LSUB$m$ and DSUB$m$) available to us for future
investigations, each of which has its own merits, and which thus
allows us more freedom in applications of the CCM to other
spin-lattice models in future.

The one slight drawback in the scheme which mitigates against our goal
of obtaining more DSUB$m$ data points, for the same computing power
than for the LSUB$m$ scheme applied to the same system, and that hence
can be used together to attain more accuracy in the extrapolations, is
the slight ``even-odd'' staggering in the data that is observed in the
DSUB$m$ results, albeit that it is somewhat reduced from the similar
stagerring in the corresponding LSUB$m$ results. We have some ideas on
how the DSUB$m$ scheme might itself be modified to reduce this
staggering and we hope to report results of these further
investigations in a future paper.

%
%
\label{last@page}
\end{document}